# Fermi level dependence of magnetism and magnetotransport in the magnetic topological insulators $Bi_2Te_3$ and $BiSbTe_3$ containing self-organized $MnBi_2Te_4$ septuple layers.

J. Sitnicka,[1] M. Konczykowski,[2] K. Sobczak,[3] P. Skupiński,[4] K. Grasza,[4] Z. Adamus,[4] A. Reszka,[4] A. Wołoś[1*]

[1] Faculty of Physics, University of Warsaw, ul. Pasteura 5, Warsaw, Poland
[2] Laboratoire des Solides Irradiés, CEA/DRF/IRAMIS, Ecole Polytechnique, CNRS, Institut Polytechnique de Paris, F- 91128 Palaiseau, France
[3] Faculty of Chemistry, CNBCh, University of Warsaw, Zwirki i Wigury 101, Warsaw, Poland
[4] Institute of Physics, PAS, Aleja Lotników 32/46, Warsaw, Poland

*corresponding author: agnieszka.wolos@fuw.edu.pl

## ABSTRACT

The magnetic coupling mechanisms underlying ferromagnetism and magnetotransport in magnetically doped topological insulators have been a central issue to gain controlled access to the magneto-topological phenomena such as quantum anomalous Hall effect and topological axion insulating state. Here, we focus on the role of bulk carriers in magnetism of the family of magnetic topological insulators, in which the host material is either $Bi_2Te_3$ or $BiSbTe_3$, containing Mn self-organized in $MnBi_2Te_4$ septuple layers. We tune the Fermi level using the electron irradiation technique and study how magnetic properties vary through the change in carrier density, the role of the irradiation defects is also discussed. Ferromagnetic resonance spectroscopy and magnetotransport measurements show no effect of the Fermi level position on the magnetic anisotropy field and the Curie temperature, respectively, excluding bulk magnetism based on a carrier-mediated process. Furthermore, the magnetotransport measurements show that the anomalous Hall effect is dominated by the intrinsic and dissipationless Berry-phase driven mechanism, with the Hall resistivity enhanced near the bottom/top of the conduction/valence band, due to the Berry curvature which is concentrated near the avoided band crossings. These results demonstrate that the anomalous Hall effect can be effectively managed, maximized, or turned off, by adjusting the Fermi level.

## I. INTRODUCTION

In recent years, there has been a continuous interest in topological magnetic materials hosting various new topological quantum states, like anomalous quantum Hall effect, topological electromagnetic effect, or axion insulator state [1–5]. One of the material systems, that has been widely utilized for possible realizations of these exotic quantum phenomena, is the topological insulator $Bi_2Te_3$ doped with Mn. The layered crystal structure of the $Bi_2Te_3$ allows the Mn dopants to enter at different positions, resulting in a diversity of magnetic properties. At a low doping level, Mn preferentially replaces the Bi sites, while the increased doping causes Mn to enter the interstitial position in the van der Waals gap [6–8]. Under the proper growth conditions, the Mn ions can form self-organized heterostructures composed of an alternating sequence of $MnBi_2Te_4$ septuple layers (SLs) and *n*-times $Bi_2Te_3$ quintuple layers (QLs). These $MnBi_2Te_4/(Bi_2Te_3)_n$ compounds belong to the family of intrinsic magnetic topological insulators (IMTI) which compared to dilute magnetic topological insulators (MTI) offer a larger magnetic gap and homogeneous magnetic ordering at the surface [9,10], making them an inviting platform for studying novel topological phenomena [11–15]. The magnetic and topological properties in these systems are highly tunable by the number *n* of QLs [16–18]. Another way of tuning the properties of magnetic topological insulators is element substitution, *e.g.*, by doping with Sb. Both magnetic and electric transport properties can be controlled by changing the proportion of Sb atoms in QLs [19–21]. Although IMTIs are structurally better organized than the diluted MTIs, the disorder effects (*e.g.,* substitution of Mn into otherwise non-magnetic $Bi_2Te_3$ QLs) have been recently recognized, strongly affecting their magnetism as well [22–24].

Despite the widespread interest in both IMTIs and diluted MTIs based on $Bi_2Te_3$, little is still known about the microscopic origin of the ferromagnetism, in particular taking into account a variety of ways of arranging magnetic dopants. Recent *ab-inito* calculations performed for diluted MTIs suggest that in the absence of free carriers the superexchange dominates magnetism, its sign being dependent on the coordination and the charge state of a magnetic dopant [25]. Remarkably, the Mn $d^5$ configuration forces antiferromagnetic superexchange which should be accompanied by the RKKY interaction due to the acceptor character of the Mn substituting Bi site. The fact that Mn-doped topological insulators have typically high concentration of free carriers, which may be either *n*- or *p*-type [6,8,14,26], indicates that their role cannot be ignored and should be carefully examined. Therefore, in this work, we conduct systematic studies to clarify the impact of the Fermi level on magnetism, in a way that excludes the influence of other factors (like different ordering of Mn) originating from ubiquitous disorder effects that can lead to different properties even in companion samples – the magnetic properties before and after irradiation can be studied on the same sample, differing only in the position of the Fermi level and the number of non-magnetic irradiation defects. We performed a series of irradiations with a high-energy electron beam on *p*-type magnetic topological insulator samples, followed by subsequent annealing steps where applicable, allowing us to trace changes in magnetization in the same sample. This method of tuning the chemical potential was previously successfully implemented for compensating charged bulk defects in $Bi_2Te_3$ and $Bi_2Se_3$ topological insulators [27], showing that with increasing the electron fluence the Fermi level can be tuned from the valence band through the band gap to the conduction band, and back after the annealing. We tested both $Bi_2Te_3$ and $BiSbTe_3$ doped with Mn. The magnetic properties were systematically investigated by ferromagnetic resonance (FMR) and magnetotransport measurements, before and after the irradiation process. Both applied measurement methods reveal that bulk magnetism does not depend on the Fermi level (magnetic anisotropy field and the Curie temperature $T_C$ remain unchanged), thus excluding the dependence on the carrier density. Moreover, magnetotransport measurements indicate that the bulk contribution to the anomalous Hall effect (AHE) is strongly correlated with the Fermi level via the Berry curvature mechanism. The dominance of this intrinsic contribution to the anomalous Hall effect is enhanced at the top of the valence band and at the bottom of the conduction band.

## II. RESULTS
### A. Structural characterization of as-grown samples

Mn-doped $Bi_2Te_3$ and $BiSbTe_3$ crystals were synthesized using the Bridgman method (for details see Methods). Several specimens were exfoliated and cut from adjacent areas of the same crystal to prepare samples for irradiation with different irradiation doses, and then to conduct various experiments. Energy dispersive x-ray analysis (EDX) shows an average Mn concentration 0.45 at. % in $Bi_2Te_3$ and 1.9 at. % in $BiSbTe_3$, respectively. In $BiSbTe_3$, the transmission electron microscopy (TEM) with EDX mapping shows homogenous distribution of Mn over the large area of the investigated specimen (Fig. 1a). It is expected that Mn mainly substitutes Bi/Sb acceptor sites which are followed by high hole concentration of the order of $5.8 \times 10^{20}$ cm$^{-3}$, reaching up to ~ $10^{21}$ cm$^{-3}$ in some areas of the crystal. The presence of a fraction of Mn in the interstitial position in van der Waals gaps cannot be ruled out yet, but they do not form a structure that can be detected in TEM. Single self-organized septuple layers of $MnBi_2Te_4$ can be detected in the $BiSbTe_3$ specimen (Fig. 1a) as well.

In $Bi_2Te_3$, TEM studies reveal the presence of numerous septuple layers, which are grouped with the preferred distance between adjacent SLs $n = 4$ QLs, forming fragments of $MnBi_2Te_4/(Bi_2Te_3)_n$ superlattice. As has been discussed earlier [22,23], in samples containing SLs Mn is not only present in the $MnBi_2Te_4$, but substitutes on Bi sites in the QLs as well. This is also visible on the EDX map in Fig. 1b as a uniform Mn distribution shown in yellow in the regions containing QLs. The $Bi_2Te_3$ crystal is *p*-type with a hole concentration $1.8 \times 10^{20}$ cm$^{-3}$.

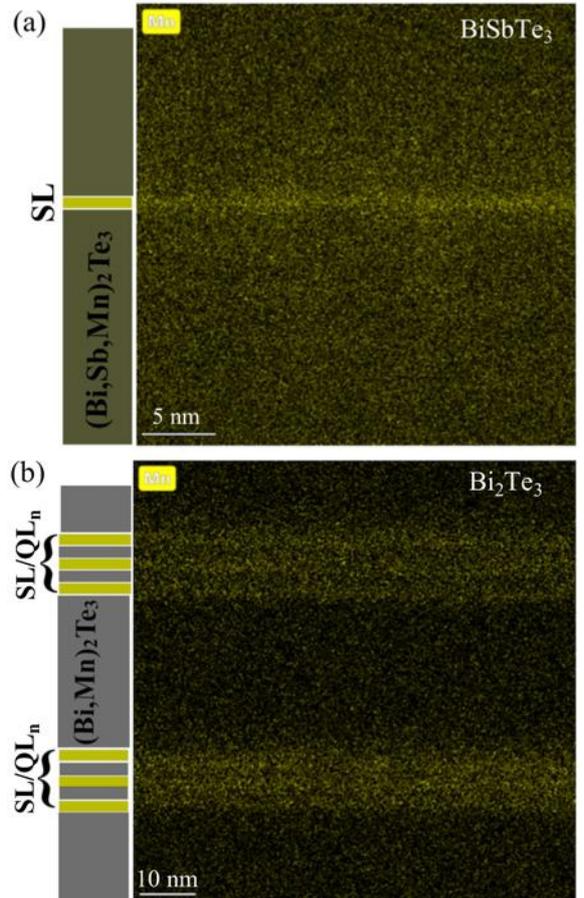

Fig. 1. EDX mapping with the distribution of Mn atoms shown in yellow for $BiSbTe_3$ in (a) and $Bi_2Te_3$ in (b). Mn is present both in the matrix $BiSbTe_3$ (a) or (QLs) of $Bi_2Te_3$ in (b) and in self-organized $MnBi_2Te_4$ SLs seen as yellow stripes.

### B. Tuning the Fermi level using an electron beam

In order to tune chemical potential and assess the effect of carrier concentration on magnetism in studied topological insulators, we used low temperature irradiation by energetic electrons already applied in a variety of TIs [14,27,28]. Exposure to the 2.5 MeV electron beam of the $Bi_2Te_3$ crystals chilled to 20 K by immersion in liquid hydrogen produces uniform spread of Frenkel (vacancy-interstitial) pairs (see details in Methods). Interstitials become mobile far below room temperature and at the room temperature efficiently either recombine with a nearby vacancy or migrate to the surface. Therefore, upon warming to transfer samples to another experimental platform, partial annihilation of the irradiation damage occurs. The remaining damage is exclusively vacancy type. The donor-type character of Bi vacancy was found to be the main contribution to the charge balance in the bulk of irradiated crystals. The other irradiation induced defects seem to play a minor role.

The Bi vacancies are much more stable than interstitials, however they also partially annihilate at room temperature. Therefore, the irradiated samples were stored in the freezer at temperature below 20 °C and transported in dry ice. Nevertheless, the time a sample stays at room temperature is a not fully controlled factor affecting the concentration of charge carriers. Thus the indicated irradiation doses discussed later in the text and shown in Table 1 serve more as labels for the tested samples and qualitatively indicate the direction of changes in the type and concentration of carriers.

The donor-type of irradiation defects and consequently following upward shift of the Fermi energy upon irradiation opens the possibility to reach the charge neutrality point (CNP) by electric compensation of preexisting defects in initially $p$-type material. Electron irradiation does not affect crystal lattice integrity and the topological Dirac surface states are preserved [27].

The penetration range of 2.5 MeV electrons has been calculated using NIST ESTAR simulator and discussed in [27]. The penetration depth for $Bi_2Te_3$ is above 2000 $\mu$m, while typical sample thickness used in this study did not exceed 60 $\mu$m, ensuring uniform profile of the irradiation over the entire sample. Nevertheless, the surface band bending is omnipresent in topological insulators, as it has been previously reported for $Bi_2Te_3$, $Bi_2Se_3$ and $Bi_2Te_2Se$ [29–31]. In irradiated and annealed samples, additionally, the effect of redistribution of Bi vacancies, which are the main type of defects affecting electronic properties, occurs under removing a sample to room temperature and further annealing. In Ref. [14] it has been argued that the surface region after thermal annealing is denuded of vacancies, which causes upward band bending and aligns the magnetic gap of the topological surface states with the conduction band states in the bulk. As a result, the quantum anomalous Hall effect is observed in the quantization window, superimposed on bulk conduction. In the thick samples studied in this work, surface effects are, however, negligible and a homogeneous distribution of irradiation defects can be assumed.

The irradiation process allows to effectively tune the Fermi level in Mn doped $Bi_2Te_3$ through the energy gap. Fig. 2a shows the resistivity of the $Bi_2Te_3$:Mn irradiated with 2.5 MeV electrons versus the irradiation dose, measured *in situ* at 20 K. The resistivity increased by two orders of magnitude at the charge neutrality point reached at about 0.34 C/cm$^2$, where the conduction converted from $p$- to $n$-type.

The pristine $Bi_2Te_3$:Mn crystal was found to be very homogeneous with the initial carrier concentration $p = 1.8 \times 10^{20}$ cm$^{-3}$. After irradiation to 1.36 C/cm$^2$ and transfer via room temperature to another experimental setup, the hole concentration reduced to $p = 3.8 \times 10^{18}$ cm$^{-3}$. The sample did not convert to the $n$-type marking clearly the offset between the *in situ* characteristics made at 20 K (Fig. 2a) and measurements performed after exposition to the room temperature (Fig. 2b). The conversion of the conductivity type after exposure to the room temperature was achieved after irradiation to the dose as high as 1.56 C/cm$^2$ resulting in electron concentration $n = 7.6 \times 10^{16}$ cm$^{-3}$. Irradiation with a higher dose, 3.26 C/cm$^2$, reveals the effects of the saturation of the electron concentration at the level of $n = 8.9 \times 10^{16}$ cm$^{-3}$, see Fig. 2b and Table 1.

Unlike $Bi_2Te_3$, in $BiSbTe_3$ doped with Mn it was not possible to achieve carrier-type conversion even at high irradiation doses, Fig. 2c. The initial hole concentration of the pristine $BiSbTe_3$ varied between $5.8 \times 10^{20}$ and $9.7 \times 10^{20}$ cm$^{-3}$. Several samples were irradiated with high doses: 1 C/cm$^2$, 2 C/cm$^2$, 2.87 C/cm$^2$, 4.57 C/cm$^2$, and 5.26 C/cm$^2$, which resulted in hole concentration: $6.2 \times 10^{20}$ cm$^{-3}$, $3.4 \times 10^{20}$ cm$^{-3}$, $3.1 \times 10^{20}$ cm$^{-3}$, $2.6 \times 10^{20}$ cm$^{-3}$, and $3.2 \times 10^{20}$ cm$^{-3}$, respectively, Table 1. The initial drop of the hole concentration (from $9.7 \times 10^{20}$ cm$^{-3}$ down to $6.2 \times 10^{20}$ cm$^{-3}$) at irradiation to 1 C/cm$^2$, *i.e.*, $3.5 \times 10^{20}$ cm$^{-3}$ per 1 C/cm$^2$, is much greater than previously found for the undoped $Bi_2Te_3$, of the order of $1 \times 10^{20}$ per 1 C/cm$^2$ [27]. This could suggest a two-channel type of conductivity and the effective character of the concentration determined from the Hall constant ($p = 1/(eR_H)$). After exposure to the highest doses, 4.57 C/cm$^2$ and 5.26 C/cm$^2$, the effective concentration showed saturation effects at the level of about $p = 3 \times 10^{20}$ cm$^{-3}$. The small scatter around the observed trend is due to the inhomogeneity of the pristine material.

The investigated Mn doped $Bi_2Te_3$ and $BiSbTe_3$ crystals were both $p$-type in a pristine state. Low temperature irradiation with 2.5 MeV electrons produced depression of hole concentration, eventually reaching conductivity type inversion (in $Bi_2Te_3$). Subsequent thermal annealing produces partial annihilation of introduced defects and/or their redistribution, reducing the compensation of bulk carriers and thus reversing the effect of irradiation [27]. It should be emphasized that this approach of carrier density modulation by irradiation and subsequent annealing allows reliably assessing the impact of the sole Fermi level position (and non-magnetic irradiation defects) on magnetism in the same sample.

Table 1. Concentration of electric charge carriers in Mn - doped $Bi_2Te_3$ and $BiSbTe_3$, measured at 2 K before and after irradiation with 2.5 MeV electrons and transfer via room temperature to another experimental setup.

| Irradiation dose (C/cm$^2$) | Carrier concentration (cm$^{-3}$) |
|---|---|
| $Bi_2Te_3$:Mn | |
| 0 | $p = 1.8 \times 10^{20}$ |
| 1.36 | $p = 3.8 \times 10^{18}$ |
| 1.56 | $n = 7.6 \times 10^{16}$ |
| 3.26 | $n = 8.9 \times 10^{16}$ |
| $BiSbTe_3$:Mn | |
| 0 | $p = (5.8 - 9.7) \times 10^{20}$ |
| 1.00 | $p = 6.2 \times 10^{20}$ |
| 2.00 | $p = 3.4 \times 10^{20}$ |
| 2.87 | $p = 3.1 \times 10^{20}$ |
| 4.57 | $p = 2.6 \times 10^{20}$ |
| 5.26 | $p = 3.2 \times 10^{20}$ |

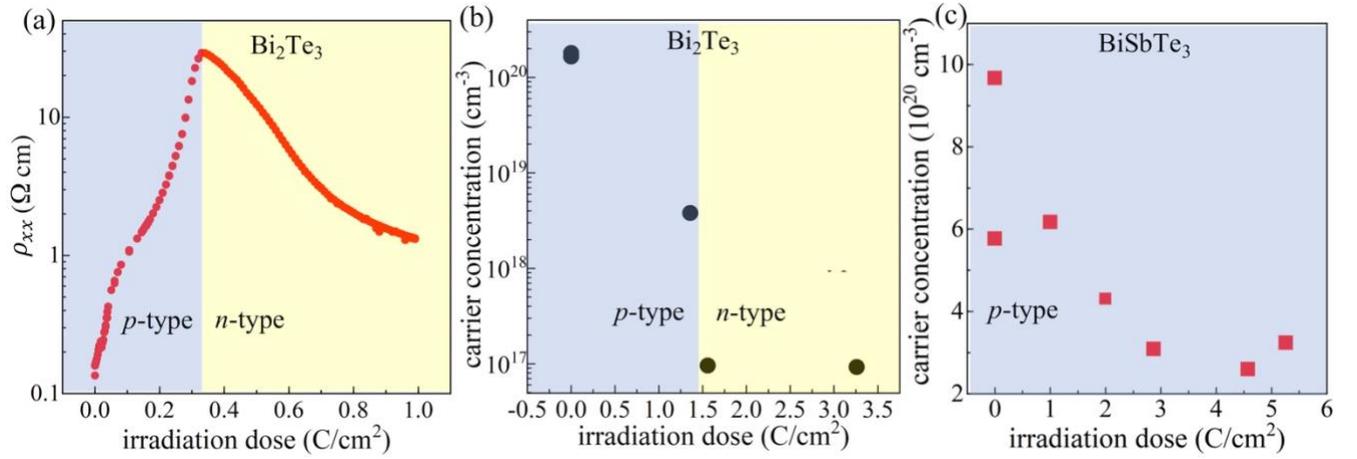

Fig. 2. (a) Longitudinal resistivity measured *in situ* at 20 K for $Bi_2Te_3$:Mn irradiated with 2.5 MeV electrons versus the irradiation dose. The resistivity shows an increase of about two orders of magnitude at the charge neutrality point at about 0.34 C/cm$^2$, where the conduction is converted from *p*- to *n*-type. Carrier concentration versus irradiation dose, measured at 2 K after transfer via room temperature to another experimental setup is shown for the $Bi_2Te_3$:Mn in (b) and for the $BiSbTe_3$:Mn in (c).

### C. The influence of the Fermi level on magnetism, ferromagnetic resonance

Ferromagnetic resonance (FMR) is a powerful technique to study the magnetic properties of all classes of magnetic media, from bulk ferromagnetic materials to nanoscale magnetic films. An FMR signal is observed due to the effect of precession of the macroscopic magnetization, as it is driven by an external magnetic field of electromagnetic radiation in the microwave range [32]. During the experiment, the frequency of the microwave field is fixed whilst an applied static magnetic field is scanned through the resonant condition. The resonance field depends on the orientation of the sample and the microwave frequency. FMR measurements can be applied to study experimentally the most important characteristic parameters of the magnetic material, in particular magnetic anisotropy constants which together determine the magnetic anisotropy of the studied system.

FMR measurements were performed on Mn-doped $Bi_2Te_3$ and $BiSbTe_3$ samples, irradiated to the dose 2.16 C/cm$^2$ for $Bi_2Te_3$ and 3.26 C/cm$^2$ for $BiSbTe_3$, respectively, before and after subsequent thermal annealing for 30 min. at temperatures, respectively, 100 - 150 °C and 200 °C. The resonance spectra were recorded at different temperatures for two orientations of the external magnetic field $H$, $H \parallel c$, and $H \perp c$, where $c$ is the crystallographic axis perpendicular to the sample surface.

Let us recall that after irradiation $Bi_2Te_3$ converted from *p*-type with hole concentration of $1.8 \times 10^{20}$ cm$^{-3}$ to *n*-type, with electron concentration of about $8-9 \times 10^{16}$ cm$^{-3}$, while $BiSbTe_3$ reduced its hole concentration from about $6 - 10 \times 10^{20}$ cm$^{-3}$ down to about $3 \times 10^{20}$ cm$^{-3}$. We have previously shown, that isochronal annealing in 30 min. intervals reduces irradiation-induced defects in $Bi_2Te_3$ when heating above about 75 °C, moving the Fermi level back to the position in pristine material when heating above 150 °C [27].

FMR spectra obtained for each irradiated sample before and after annealing are shown in Fig. 3a. Single resonance line allows treating precession of the magnetic moments with uniform resonance mode approximation, despite the complex magnetic structure of samples composed of both SLs and Mn-doped QLs [22]. Remarkably, no significant change in the character of the spectra was observed upon annealing, clearly indicating that magnetism in studied samples is independent of the position of the Fermi level. This observation is fully consistent with an earlier report on *n*-type $MnBi_2Te_4/(Bi_2Te_3)_n$ superlattices, where the Curie temperature was shown to be independent of the free electron concentration (see Supplementary Information in [14]).

Figures 3b and 3c show the position of the resonance line for $Bi_2Te_3$ and $BiSbTe_3$, respectively, in two sample orientations relative to the external magnetic field, ***H*** ∥ ***c***, and ***H*** ⊥ ***c***, measured versus temperature. The FMR occurs at a lower magnetic field for ***H*** applied parallel to the ***c*** axis than for ***H*** applied perpendicular to it, Figure 3 a-c, due to large and positive magnetocrystalline anisotropy which dominates over the demagnetization anisotropy [22] and establishes the easy axis out of-plane (along the ***c*** direction). The anisotropy of the ferromagnetic resonance decreases while increasing temperature and the resonance takes its paramagnetic position above the Curie temperature with *g* - factor equal to 1.9 ± 0.2 alike for both $Bi_2Te_3$ and $BiSbTe_3$. The obtained paramagnetic *g*-factor values are characteristic for the $Mn^{2+}$ ($d^5$) configuration [7,22].

In crystals with dominating axial anisotropy, the difference between the resonance field for ***H*** applied parallel to the ***c*** axis, $H_{res}(0°)$, and perpendicular to it, $H_{res}(90°)$, is three times the anisotropy field (see the spin Hamiltonian in ref. [22]):

$$H_{res}(0°) - H_{res}(90°) = 3\, H^A. \quad (1)$$

The magnetic anisotropy field, $H^A$, determined from the above equation is shown in Figure 3d as a function of temperature. The magnetic anisotropy field is an effective parameter describing the energy needed to rotate the magnetization direction from the easy axis to the hard axis, thus characterizing the magnetic properties of a given material. It can be seen from Fig. 3d that the $H^A$ over the entire temperature range is nearly unchanged by the annealing after irradiation, which returns the Fermi level back to its original position in the valence band. The $H^A$ decreases with increasing temperature, however, it vanishes gradually with a long tail extending above the $T_C$ due to the magnetic correlations that survive to the paramagnetic regime [22]. Therefore, due to the lack of clear features in $H^A(T)$ to determine the $T_C$, this parameter will be discussed separately in the next chapter.

The FMR experiment shows that the annealing after irradiation, thus returning the Fermi level back to the original position in the valence band, does not significantly affect the magnetic anisotropy field, which indicates that the position of the Fermi level has no effect on magnetic properties. We can extend this conclusion by stating that irradiation defects also do not affect the magnetism. This finding is consistent with the recent study on disordered *n* - type $MnBi_2Te_4/(Bi_2Te_3)_n$ showing no effect of carrier concentration change on the $T_C$ [14]. This is in contrast to the expected RKKY interaction [25], however in these materials, we have demonstrated earlier that Mn is not randomly distributed but planarly self-organized in SLs. There is a ferromagnetic coupling between SLs and the residual population of Mn in QLs [22], which indicates the qualitative difference between IMTIs containing SLs and widely discussed diluted MTIs [25].

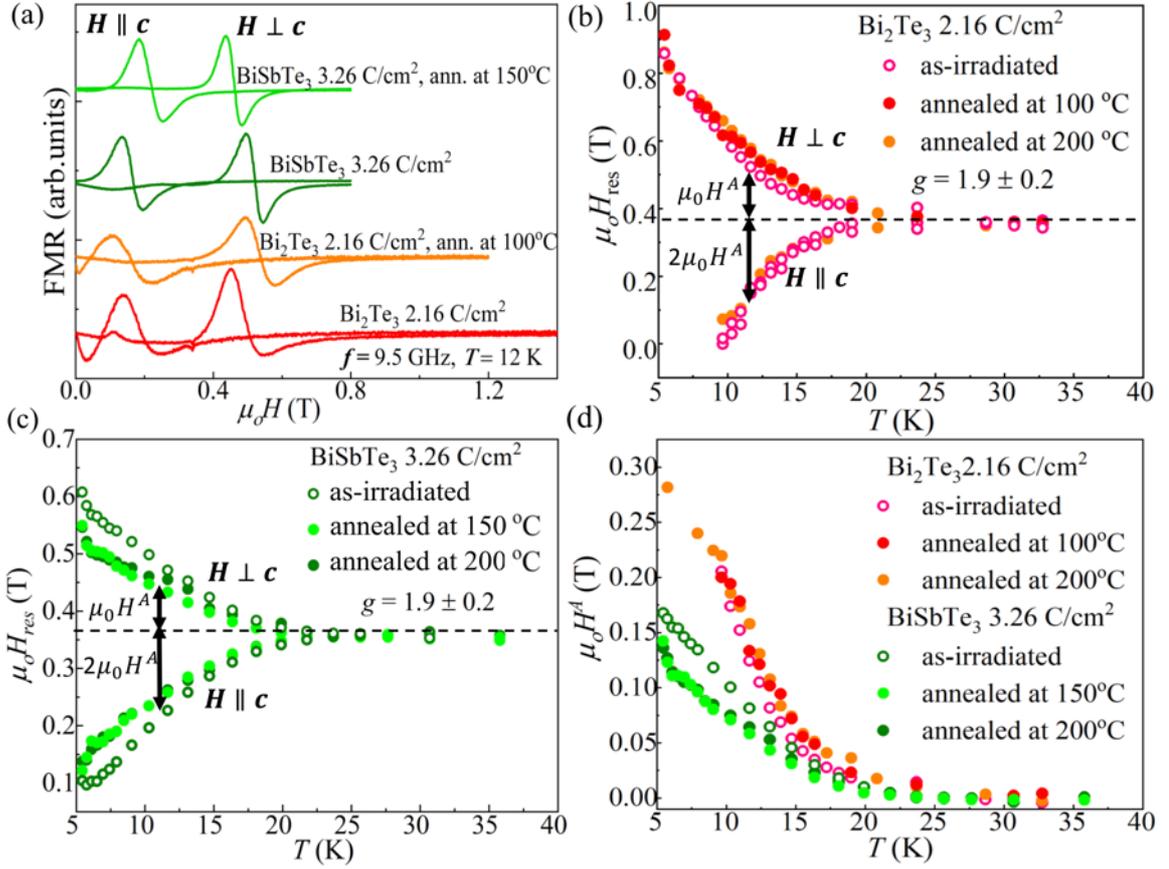

Fig. 3. X-band FMR spectra of BiSbTe$_3$ and Bi$_2$Te$_3$ doped with Mn irradiated to the dose 3.26 C/cm$^2$ for BiSbTe$_3$ and 2.16 C/cm$^2$ for Bi$_2$Te$_3$, respectively, and subsequently annealed measured for magnetic field applied parallel and perpendicular to the *c* axis are shown in (a). The resonance field $\mu_0 H_{res}$ for $H \parallel c$ and $H \perp c$, respectively, determined versus the temperature for irradiated samples before and after annealing are shown for Bi$_2$Te$_3$ in panel (b) and for BiSbTe$_3$ in panel (c). Magnetic anisotropy field $\mu_0 H^A$ determined from Equation (1) versus the temperature for Bi$_2$Te$_3$ and BiSbTe$_3$ samples as-irradiated and subsequently annealed is shown in (d). The magnetic anisotropy field is unaffected by irradiation and subsequent annealing indicating the insensitivity of the magnetic properties to the Fermi level.

### D. The influence of the Fermi level on magnetism, magnetotransport measurements

Magnetotransport studies aimed at assessing independently the influence of the Fermi level on the another parameter characterizing properties of magnetic material, *i.e.*, Curie temperature $T_C$, were carried out on Bi$_2$Te$_3$ pristine and irradiated to 1.56 C/cm$^2$ and 3.26 C/cm$^2$ as well as BiSbTe$_3$ pristine and irradiated to 2.87 C/cm$^2$ and further to the cumulative dose 4.57 C/cm$^2$, respectively. Magnetotransport measurements were performed in a van der Pauw geometry with the current flowing in the *ab*-plane. Hall data measured at 4.2 K in magnetic field up to 4 T are shown in the Appendix. Hall measurements reveal heavy *p* - type doping induced during synthesis in both pristine samples, resulting in Hall concentration $1.8 \times 10^{20}$ cm$^{-3}$ in Bi$_2$Te$_3$ and 6-10 $\times 10^{20}$ cm$^{-3}$ in BiSbTe$_3$, as discussed earlier. Irradiation of Bi$_2$Te$_3$ leads to the type conversion from *p*- to *n*-type (Fig. 2b), with electron concentration of the order of 8-9 $\times 10^{16}$ cm$^{-3}$ at high irradiation doses while irradiation of BiSbTe$_3$ reduces only the hole concentration down to about $3 \times 10^{20}$ cm$^{-3}$ at high irradiation doses (Fig. 2c).

Interestingly, pristine Bi$_2$Te$_3$ shows no anomalous Hall effect (Fig. 4a) despite the clear detection of the ferromagnetic resonance (Fig. 3). Simultaneously, magnetization measurements performed using a Hall probe applied to the sample surface indicate hysteresis loops of a ferromagnetic material (Fig. 4b). The saturation magnetization plotted versus temperature and fitted with $M_{sat} \sim \left(1 - \frac{T}{T_C}\right)^\alpha$ revealed Curie temperature $T_C$ equal to 13 K, Fig. 4c, which is simultaneously the critical temperature for ferromagnetic ordering in a single MnBi$_2$Te$_4$ layer and is characteristic for highly ordered ferromagnetic MnBi$_2$Te$_4$/(Bi$_2$Te$_3$)$_n$ samples [22].

While the resistivity hysteresis loops discussed below (appearing after irradiation in Fig. 4a) have nearly rectangular shape, the magnetization measured by the Hall probe put on a sample surface shows discontinuities in the slopes versus applied magnetic field, indicating abrupt transitions between different magnetic regimes. Inset to Fig. 4b shows separation of these regimes in the phase diagram. Due to the surface character of the measurement made with the Hall probe, it is reasonable to assume that the observed effect is related to near-surface areas. The nature of the different magnetic regimes in ferromagnetic MnBi$_2$Te$_4$/Bi$_2$Te$_3$ (with Bi$_{Mn}$ antisites in Bi$_2$Te$_3$ QLs) is not clear. So far, various spin textures have been identified and found responsible for the observed features in the AHE in antiferromagnetic $MnBi_2Te_4$ without intercalating $Bi_2Te_3$ [33,34]. In the case of $MnBi_2Te_4/Bi_2Te_3$, this problem requires further microscopic and theoretical studies, in particular in the context of obtaining QAHE, which is crucially dependent on the spin texture of the subsurface regions.

After electron irradiation and change of the conductivity to the *n*-type, the anomalous Hall effect (AHE) appeared in $Bi_2Te_3$ with clear hysteresis loops in transverse Hall resistivity $\rho_{xy}$ (Fig. 4a and Fig. S1a in the Appendix). Note, that pristine material in Fig. 4a is the same sample as irradiated to 1.56 C/cm$^2$. The temperature dependence of the anomalous Hall resistivity at magnetic saturation $\rho_{sat}^{AHE}$ is plotted in Fig. 4c together with saturation magnetization from the Hall probe for pristine sample shown for comparison. Clearly, all the curves in Fig. 4c indicate the same $T_C = 13$ K, consistent with the conclusion from the FMR measurements that irradiation, both changing the position of the Fermi level and introducing irradiation defects, does not affect the magnetism in studied samples. The same conclusion applies to $BiSbTe_3$, for which AHE is well pronounced in both pristine and irradiated samples (Fig. 5a and Fig. S1b in the Appendix) and the dependence of the anomalous Hall resistivity at magnetic saturation $\rho_{sat}^{AHE}$ on temperature indicates in all samples the same $T_C = 16$ K (Fig. 5b).

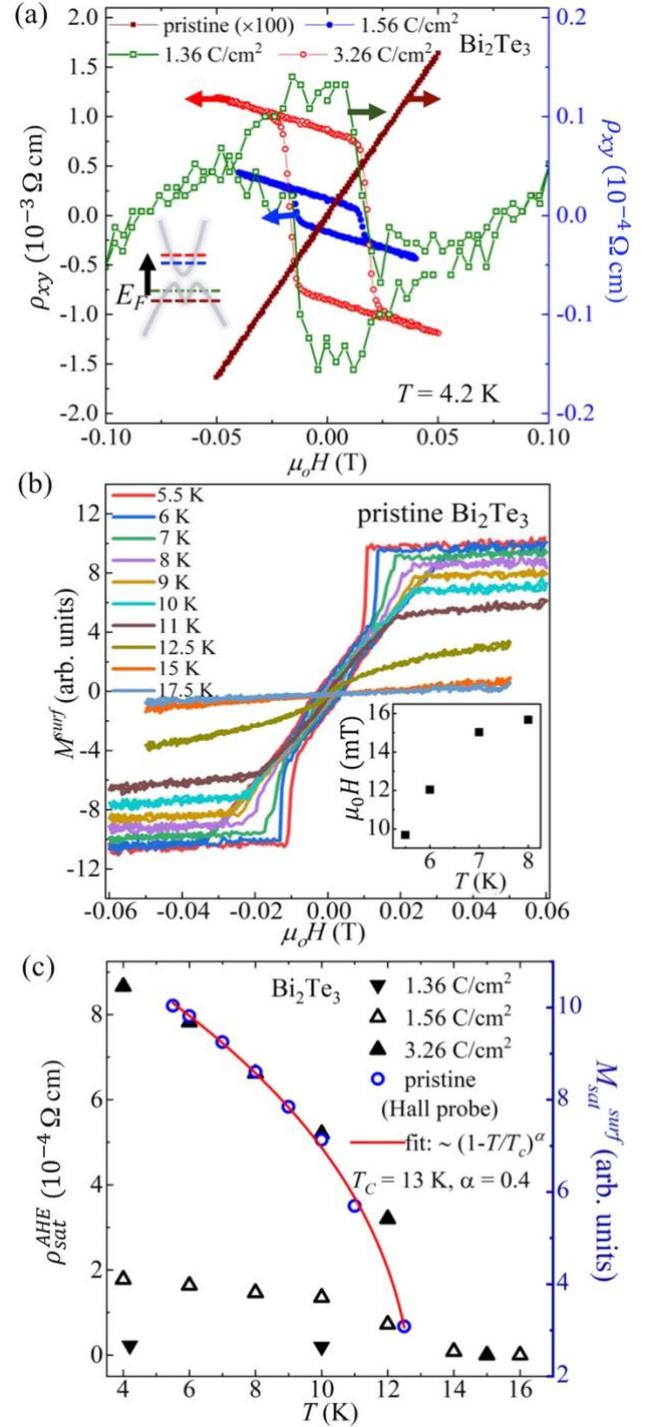

Fig. 4. (a) Transverse Hall resistivity $\rho_{xy}$ vs. out-of-plane magnetic field $H$ for pristine and irradiated $Bi_2Te_3$ measured at 4.2 K. Pristine $\rho_{xy}$ data is multiplied by 100 for improved presentation. Inset: schematic of $E_F$ position in band structure by irradiation dose (color-coded). The arrow shows $E_F$ changes with increasing irradiation dose. (b) Surface magnetization $M_{surf}$ (difference between magnetic induction measured by Hall sensor placed on the sample surface and applied $H$) of pristine $Bi_2Te_3$:Mn measured using a Hall probe at different $T$. The $M_{surf}$ shows discontinuities in the slopes versus applied $H$, indicating abrupt transitions between different magnetic regimes. Inset shows phase diagram separation of these regimes. (c) $\rho^{AHE}$ at magnetic saturation vs. $T$ for $Bi_2Te_3$:Mn irradiated to dose 1.56 C.cm$^2$ ($n = 7.6 \times 10^{16}$), and 3.26 C/cm$^2$ ($n = 8.9 \times 10^{16}$), respectively, is shown together with $M_{surf}$ at saturation obtained from Hall probe for pristine material. All curves show $T_C \sim 13$ K independent on the $E_F$. The red solid line shows theoretical fit.

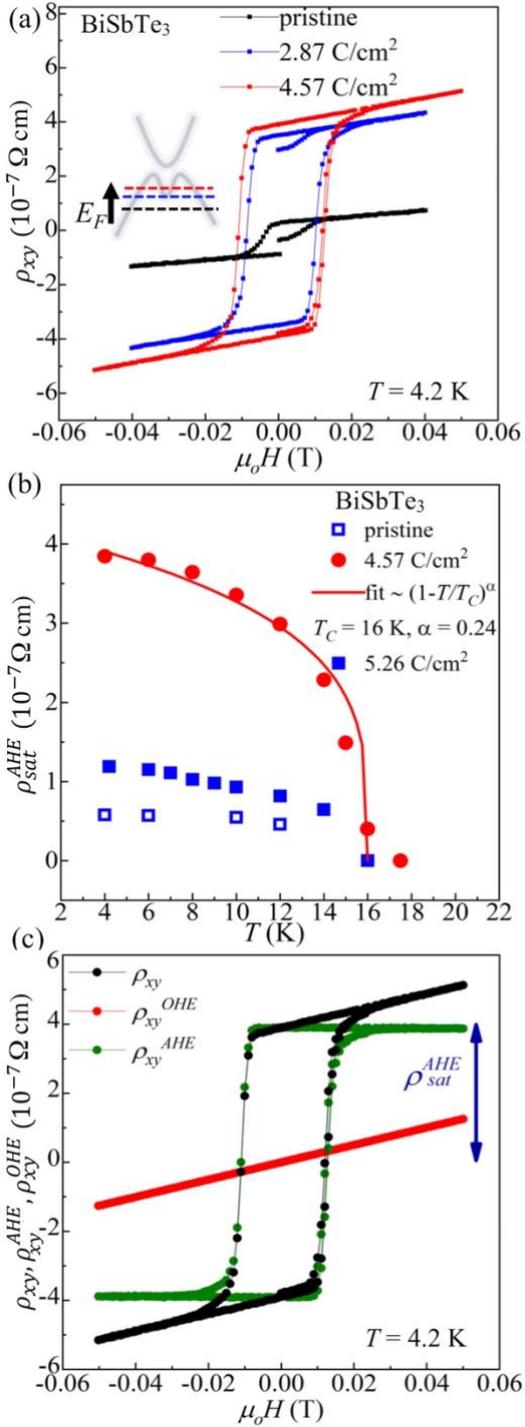

Fig. 5. (a) Transverse Hall resistivity $\rho_{xy}$ vs. out-of-plane magnetic field for pristine and irradiated BiSbTe$_3$, measured at 4.2 K. Inset shows a schematic demonstration of the Fermi level $E_F$ position in the band structure for a given dose of irradiation (color assignment to irradiation dose). The arrow indicates the direction of changes in the position of $E_F$ with increasing irradiation dose. (b) $\rho^{AHE}$ at magnetic saturation vs. $T$ for BiSbTe$_3$:Mn pristine and irradiated to dose 4.57 C/cm$^2$ ($p = 2.6 \times 10^{20}$) and 5.26 C/cm$^2$ ($p = 3.2 \times 10^{20}$), respectively. All curves show $T_C \sim 16$ K independent on the Fermi level. (c) $\rho_{xy}^{OHE}$ (red line) and $\rho_{xy}^{AHE}$ (green line) have been separated from $\rho_{xy}$ (black line). The value of $\rho^{AHE}$ at magnetic saturation is marked with a blue arrow.

### E. The influence of the Fermi level on the anomalous Hall effect

The anomalous Hall effect is a fundamental transport phenomenon that occurs in systems that do not have time-reversal symmetry. MTIs with strong spin-orbit coupling are promising platforms for realizing both the quantized version of the anomalous Hall effect [1–3,14] and its classic form [35]. In the classic AHE, three regimes are typically distinguished and connected to different microscopic origins, clean, intermediate, and dirty [36]. In the clean limit, $\sigma_{xy}^{AHE}$ conductivity is dominated by the extrinsic skew scattering mechanism which is proportional to the transport lifetime $\tau$, yielding $\sigma_{xy}^{AHE} \propto \sigma_{xx}$. In the intermediate regime, the skew scattering mechanism is suppressed by increased impurity scattering and the intrinsic Berry-phase driven mechanism dominates [36–38]. In this case, $\sigma_{xy}^{AHE}$ becomes insensitive to the scattering time thus independent of $\sigma_{xx}$. In the strongly disordered or hopping (dirty) regime $\sigma_{xy}^{AHE}$ takes the dependence on $\sigma_{xx}$ in the form $\sigma_{xy}^{AHE} \propto \sigma_{xx}^{\beta}$, with $\beta = 1.6 - 1.8$. The microscopic origin underlying this scaling is not yet known but it might be related to the broadening of the electronic spectrum [36]. The dirty regime is observed in materials exhibiting variable range hopping transport. The investigated Bi$_2$Te$_3$ and BiSbTe$_3$ doped with Mn fall in the intermediate regime with the dissipationless topological Berry curvature mechanism governing the AHE. The extreme regimes (clean and dirty) can be excluded with the mobility of our samples between 118 and 1000 cm$^2$/(Vs) (Fig. 6). Although electron irradiation increases the number of defects, which leads to a decrease in mobility (Fig. 6), the mobility after the irradiation is still far from typical values for the hopping transport, which is often of the order of unity [39]. This is a unique feature of our experimental approach, we tune the Fermi level in a huge range, over the band gap, while only slightly changing the mobility.

The Hall resistivity $\rho_{xy}$ in a ferromagnet is expressed by the empirical relation: [36]

$$\rho_{xy} = \rho_{xy}^{OHE} + \rho_{xy}^{AHE}, \quad (2)$$

where the $\rho_{xy}^{OHE} = R_o H_z$ is the ordinary Hall effect (OHE) resistivity proportional to the magnetic field perpendicular to the sample surface $H_z$, described by the normal Hall parameter, $R_o$. The second term in (2) represents AHE contribution, which is supposed to scale with sample magnetization $M_z$ and is characterized by the anomalous Hall coefficient $R_A$, $\rho_{xy}^{AHE} = M_z R_A$. Indeed, it has been shown that in a topological antiferromagnet MnBi$_2$Te$_4$, $\rho_{xy}^{AHE}$ scales linear or superlinear with magnetization at fixed Fermi level [33]. Figure 5c illustrates schematically the contribution of the anomalous Hall effect $\rho_{xy}^{AHE}$ and ordinary Hall effect $\rho_{xy}^{OHE}$ to the measured total transverse Hall resistivity $\rho_{xy}$.

For the intrinsic Berry-phase mechanism, the anomalous Hall coefficient $R_A$ is proportional to the square of the longitudinal resistivity $\rho_{xx}$:

$$R_A = a\rho_{xx}^2, \quad (3)$$

where $a$ is a parameter proportional to the Berry curvature which is derived from the Berry phase, or equivalently:
$$\sigma_{xy}^{AHE} = aM_z \quad (4)$$
after inversion of the resistivity tensor ($\sigma_{xy} \approx \rho_{xy}/\rho_{xx}^2$).

Figure 7 shows $\sigma_{xy}^{AHE}$, obtained by subtracting ordinary Hall component, versus $\sigma_{xx}$ measured at different temperatures in the ferromagnetic phase for $Bi_2Te_3$ irradiated with doses 1.56 C/cm$^2$ and 3.26 C/cm$^2$ and $BiSbTe_3$ pristine and irradiated with dose 4.57 C/cm$^2$, respectively. Since the Curie temperature is ~ 13 K for $Bi_2Te_3$ (Fig. 4c) and ~ 16 K for $BiSbTe_3$ (Fig. 5b), the temperature range included in the analysis was $T \leq 10$ K for $Bi_2Te_3$ and $T \leq 12$ K for $BiSbTe_3$, to assure nearly uniform magnetization. The variability range of $\sigma_{xx}$ is unfortunately small, limited by the small $T_c$. However, the relative change of the AHE conductivity $\Delta\sigma_{xy}^{AHE}$ with respect to the change of longitudinal conductivity $\Delta\sigma_{xx}$ varies between 0.01 % and 10 %, therefore it is not in contrast to the assumed earlier intrinsic mechanism related to the Berry phase, for which $\sigma_{xy}^{AHE}$ should be independent on $\sigma_{xx}$. This remains valid both before and after irradiation, despite the introduction of the irradiation defects, indicating that irradiation does not introduce enough disorder to alter the mechanism of the AHE (see in particular Fig. 7b with the highest change of $\sigma_{xx}$).

It is also necessary to comment at this point, that the side-jump mechanism for AHE, not discussed before, is another mechanism that predicts quadratic dependence $R_A \propto \rho_{xx}^2$ (or Hall conductivity independent of the longitudinal conductivity) which is the same scaling law as for the intrinsic mechanism. It is thus experimentally difficult to distinguish these two contributions, however, the side-jump mechanism is generally expected to be smaller than the intrinsic contribution [40]. Moreover, as the intrinsic AHE can be evaluated quantitatively from the band structure calculations even for relatively complex materials, the practical approach is to check whether the calculation explains the observation. If it is the case the intrinsic contribution seems to dominate the AHE in the studied system [36].

Indeed, the theory explains our experiment, as the Berry curvature is resonantly enhanced near avoided crossings in the energy band structure. *Ab-initio* calculations for the general 3D model of TIs [14,38] show that when the Fermi energy lies in the band gap the $\sigma_{xy}^{AHE}$ is zero. The $\sigma_{xy}^{AHE}$ becomes resonantly enhanced when the Fermi level enters the conduction band/or the valence band (avoiding crossings of band inversions) and then quickly approaches zero when the Fermi level is moved above the bottom of the conduction band/or below the top of the valence band. We can trace this relationship for the irradiated $Bi_2Te_3$ since with the irradiation experiment we cover both *p*- and *n*-type of conductivity. Moreover, in the irradiation experiment, we tune the Fermi level while magnetization is nearly fixed, as confirmed by the FMR measurements in Fig. 3, thus while changing Fermi energy we effectively probe changes in the Berry curvature, parameter $a$ in Eq. (4). Figure 8 shows $\sigma_{xy}^{AHE}$ versus Fermi energy calculated from the carrier density $n$ or $p$, respectively, using scaling proposed by Köhler [41,42]. $\sigma_{xy}^{AHE}$ is zero when the Fermi level is deeply in the valence band and increases when the Fermi level moves towards the top of the valence band. In the *p*-type sample with hole concentration $1.8 \times 10^{20}$ cm$^{-3}$, which according to the scaling for the ideal $Bi_2Te_3$ corresponds to the Fermi energy about $E_F = 25$ meV below the top of the valence band [42] there is indeed no AHE. When the Fermi level is moved towards the top of the valence band, the AHE appears for $p = 3.8 \times 10^{18}$ cm$^{-3}$ ($E_F = 22$ meV below the top of the valence band). After the conversion to *n*-type one can observe a resonant enhancement of $\sigma_{xy}^{AHE}$ close to the bottom of the conduction band for $n = 7.6 \times 10^{16}$ cm$^{-3}$ ($E_F = 2.4$ meV above the bottom of the conduction band) and $8.9 \times 10^{16}$ cm$^{-3}$ ($E_F = 2.6$ meV above the bottom of the conduction band), consistent with theoretical considerations. For $BiSbTe_3$, the range of concentration variability is small, however, the increase of the $\sigma_{xy}^{AHE}$ from 4.1 ($\Omega$ cm)$^{-1}$ to 8.8 ($\Omega$ cm)$^{-1}$ when tuning the hole concentration from $5.8 \times 10^{20}$ cm$^{-3}$ to $2.6 \times 10^{20}$ cm$^{-3}$, thus moving the Fermi level towards the top of the valence band, is also consistent with the Berry-phase mechanism. This non-linear dependence of the ordinary AHE on the Fermi level allowed to observe the quantized version of the anomalous Hall effect (QAHE) in *n*-type $MnBi_2Te_4/(Bi_2Te_3)_n$ superlattices through a quantization window with suppressed contribution from the bulk AHE [14]. In pristine samples, QAHE was covered by the bulk contribution of the AHE. Irradiation with the electron beam and subsequent annealing allowed to fully suppress AHE and reveal the quantization. It is worth emphasizing that this is the first communication experimentally showing the non-monotonic dependence of AHE on the position of the Fermi level, which requires further theoretical research and more experimental studies. However, it is a strong starting point for developing this issue.

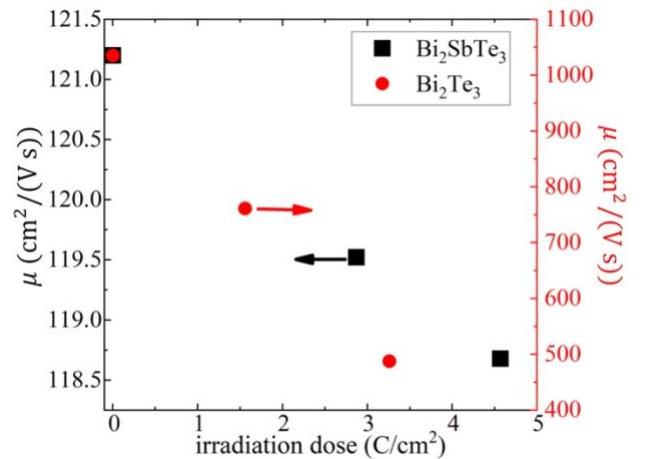

Fig. 6. The mobility as a function of the irradiation dose for the $Bi_2Te_3$ and the $BiSbTe_3$ slightly decreases after the irradiation.

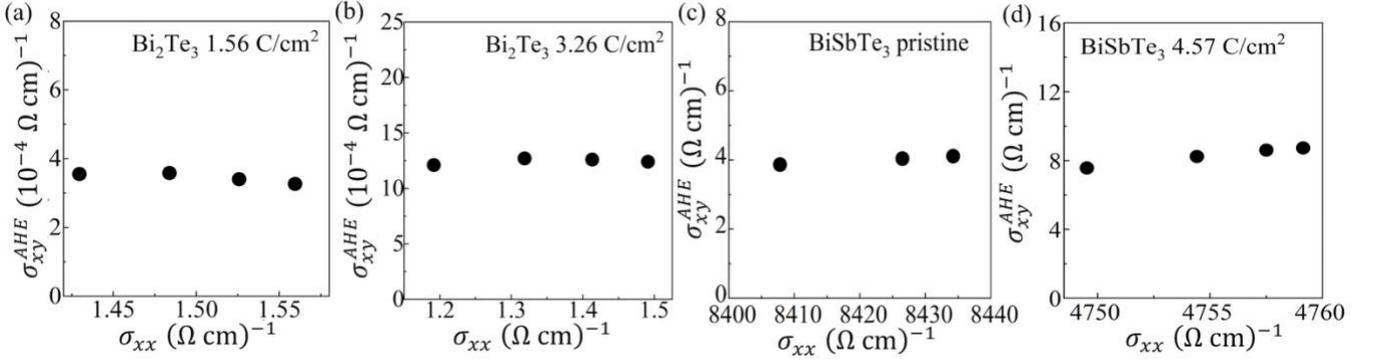

Fig. 7. $\sigma_{xy}^{AHE}$ vs. $\sigma_{xx}$ for the Bi$_2$Te$_3$:Mn sample irradiated with dose 1.56 C/cm$^2$ in (a) and 3.26 C/cm$^2$ in (b). $\sigma_{xy}^{AHE}$ does not depend on $\sigma_{xx}$ consistent with the dominant contribution from the intrinsic mechanism to AHE. The same for the BiSbTe$_3$:Mn irradiated with a dose of pristine in (c) and 4.57 C/cm$^2$ in (d). $\sigma_{xy}^{AHE}$ is also nearly independent of $\sigma_{xx}$.

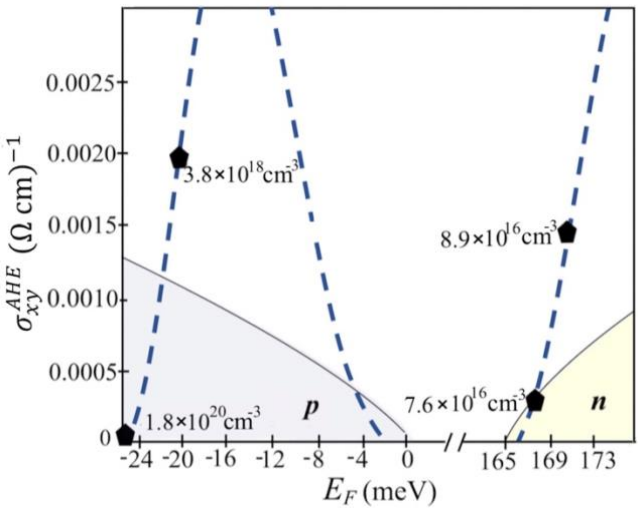

Fig. 8. $\sigma_{xy}^{AHE}$ at magnetic saturation versus the Fermi energy $E_F$, calculated from the Hall carrier density using the scaling from Refs. [41,42] for the Bi$_2$Te$_3$ containing SLs. Corresponding carrier density is indicated next to each point. Since the magnetization ($M_z$) does not vary with the $E_F$, the $\sigma_{xy}^{AHE}$ dependence on the Berry-phase (parameter $a$) is here effectively probed, $\sigma_{xy}^{AHE} = aM_z$ (Eq. 4). The blue dashed curves are a guide for the eye showing schematically the enhancement of the Berry-phase near the top of the valence band (marked in blue) or the bottom of the conduction band (marked in yellow). The bulk energy gap was assumed to be 165 meV after [43].

## III. CONCLUSIONS

We investigated the dominant contribution to the bulk AHE in Mn-doped Bi$_2$Te$_3$ and BiSbTe$_3$ containing MnBi$_2$Te$_4$ septuple layers. The main AHE comes from the intrinsic Berry-phase driven mechanism, which scales non-linear with the position of the Fermi level, consistent with recent findings [14] that the non-quantized anomalous Hall conductivity quickly approaches zero when the Fermi level moves from the top into the valence band (or from the bottom into the conduction band). This is due to the property of the Berry curvature, which is concentrated near avoided band crossings. Our research is thus the first experimental study to show the non-monotonic dependence of AHE on the position of the Fermi level. At the same time, we demonstrate here that maximizing or canceling the bulk AHE contribution is possible by tuning the Fermi level. AHE should therefore be used with caution as a method of assessing magnetic properties since it disappears away from the bottom/top of the conduction/valence band.

We investigated also the influence of the Fermi level on magnetism. Our FMR and magnetotransport measurements revealed that both the magnetic anisotropy constants and bulk $T_C$ are independent on the Fermi level, thus excluding a carrier-mediated mechanism in ferromagnetic ordering. Inconsistency with theories developed for DMTIs indicates that magnetic topological insulators with Mn ordered in SLs constitute distinct magnetic systems and their magnetism still requires in-depth studies.

## METHODS

*Sample growth*

Samples for the studies were grown by horizontal variants of the Bridgman method in the Institute of Physics, Polish Academy of Sciences. The elements used in the growth processes were high-purity (99.999 %) bismuth (Bi), tellurium (Te), antimuonium (Sb), and manganese (Mn). The materials were synthesized in quartz ampules and sealed under a vacuum (about 10$^{-4}$ Pa). Then the tube was placed in the furnace containing two heating zones. In the first step, the ampules were heated up to a temperature of about 1053 K for 48 hours to synthesize the compound and homogenize the melt. Then the ampules were cooled down to the temperature of 903 K. Next, the ampules were pulled through the temperature gradient equal to 10 K per cm at a rate of 1 mm per hour. The single crystals obtained this way had average dimensions of 50 mm × 10 mm × 8 mm.

*Structural characterization*

Structural studies were performed using transmission electron microscopy (TEM). TEM investigations were carried out by the FEI Talos F200X microscope operated at 200 kV. EDX spectroscopy using a Super-X system with four SDDs was applied to the detection of differences in local chemical composition. The samples for TEM investigations were cut along the *c*-axis, in the [11$\underline{2}$0] orientation, using a focused ion beam method.

*Electron irradiation*

Electron irradiations were carried out on "SIRIUS" facility operated by Laboratoire de Physique des Solides at École Polytechnique, Palaiseau, France. It is based on NEC Pelletron-type electrostatic accelerator coupled with a low-temperature irradiation chamber filled with liquid hydrogen fed from a close-cycle refrigerator. All irradiations were performed with samples kept at 20 K, below the mobility threshold of all irradiation induced defects (vacancies and interstitials). The size of the 2.5 MeV electron beam spot was reduced to 5 mm by a circular diaphragm aperture, with uniformity of the beam current ensured by a constant beam sweep in x and y-directions at two non-commensurate frequencies. Monitoring of the beam current density, typically of 2 μA on a 0.2 cm$^2$ surface, was performed by integration of current collected on a Faraday cage placed behind the sample chamber. Beam current densities, limited to 10 μC cm$^{-2}$ s$^{-1}$ by the cooling rate, allowed modifications of carrier concentration on the order of 10$^{20}$ cm$^{-3}$ within typical allocated beam time.

*Ferromagnetic Resonance Spectroscopy (FMR)*

The FMR measurements were performed using *Bruker ELEXSYS-E580 Electron Paramagnetic Resonance spectrometer* operating at the X-band microwave frequency (9.5 GHz). Due to the use of magnetic field modulation and the lock-in technique the resonance signal represents the field derivative of the absorbed microwave power. This device is equipped with a continuous flow helium cryostat covering the temperature range $T = 3 - 300$ K and a goniometer allowing control of the angle between the sample normal and applied external magnetic field.

In the FMR experiment, the magnetization of a ferromagnetic sample performs precession around the static magnetic field with frequency in the microwave range. Ferromagnetic resonance occurs at a fixed microwave frequency while sweeping an external magnetic field and is detected by a maximum of microwave absorption. This technique is applied widely to study magnetic anisotropy and exchange coupling in ultrathin films and multilayered materials [16,44–47].

*Transport measurements*

The irradiation system was set up for *in-situ* transport measurements that could be monitored as a function of electron dose in real-time. Crystals used in the irradiation experiments were 20 μm, 20 μm, 16 μm, and 64 μm thick with spark-weld electrical contacts placed in van der Pauw contact configuration. *Ex-situ* electrical transport measurements were performed in a van der Pauw geometry with the current flowing in the *ab*-plane with magnetic field up to 4 T. Annealing experiments up to 200 °C were performed in a Vacuum Drying Oven VT 6025.

## ACKNOWLEDGMENTS

This paper was supported by the Polish National Science Center grant 2016/21/B/ST3/02565 and the Agence Nationale de la Recherche project "DYNTOP" ANR-22-CE30-0026-01. The authors acknowledge support from the EMIR&A French network (FR CNRS 3618) on the "SIRIUS" platform under Proposal No. 18-5155, 16-9262. We thank the whole SIRIUS team, O. Cavani, R. Grasset, for operating the electron irradiation facility.

## APPENDIX

Figure S1 shows Hall resistivity $\rho_{xy}$ vs. out-of-plane magnetic field up to 4 T for pristine and irradiated Bi$_2$Te$_3$ and BiSbTe$_3$ containing self-organized MnBi$_2$Te$_4$ SLs, respectively. The Hall resistivity is linear at high magnetic field, which allows to subtract the contribution of the ordinary Hall effect and obtain the anomalous Hall effect according to Eq. 2. The Hall concentration $p$ (or $n$) was then calculated from the Hall coefficient $|R_0| = \frac{1}{pe}$ (or $|R_0| = \frac{1}{ne}$ for the negative slope), where $e$ is electron charge. The slope changes from positive in pristine Bi$_2$Te$_3$ to negative in Bi$_2$Te$_3$ irradiated to high doses, indicating the change of the conductivity type upon irradiation. BiSbTe$_3$ remains $p$ – type before and after irradiation, however the change in the inclination of $\rho_{xy}$ indicates shift of the Fermi level towards the top of the valence band. Longitudinal and Hall resistivities in non-magnetic Bi$_2$Te$_3$ have been earlier studied in Ref. [27]. A $p$ – type to $n$ – type conversion was also achieved after electron irradiation, which was clearly seen in the Hall resistance reversing its slope and thee Hall coefficient changing its sign.

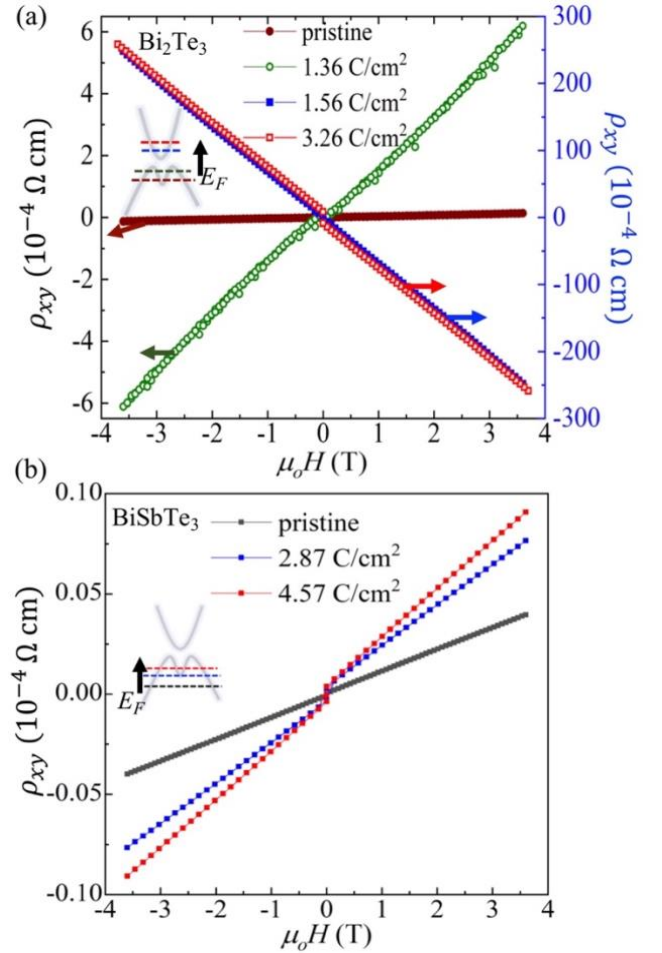

Fig. S1. Transverse Hall resistivity $\rho_{xy}$ vs. out-of-plane magnetic field measured at 4.2 K for pristine and irradiated Bi$_2$Te$_3$ in (a) and pristine and irradiated BiSbTe$_3$ in (b), both containing SLs. Insets in both panels show schematic diagrams of the Fermi level position in the band structure for a given irradiation dose (the color is assigned to the appropriate irradiation dose). The black arrows indicate the direction of changes in the position of the $E_F$ with increasing irradiation dose.

**REFERENCES:**

[1] R. Yu, W. Zhang, H.-J. Zhang, S.-C. Zhang, X. Dai, and Z. Fang, Quantized anomalous Hall effect in magnetic topological insulators, Science **329**, 5987 (2010).

[2] K. Nomura and N. Nagaosa, Surface-Quantized Anomalous Hall Current and the Magnetoelectric Effect in Magnetically Disordered Topological Insulators, Phys. Rev. Lett. **106**, 166802 (2011).

[3] J. Wang, B. Lian, and S. C. Zhang, Quantum anomalous Hall effect in magnetic topological insulators, Phys. Scr. **T164**, 014003 (2015).

[4] X. L. Qi, R. Li, J. Zang, and S. C. Zhang, Inducing a Magnetic Monopole with Topological Surface States, Science **323**, 5918 (2009).

[5] R. Li, J. Wang, X. L. Qi, and S. C. Zhang, Dynamical axion field in topological magnetic insulators, Nat. Phys **6**, 284 (2010).

[6] Y. S. Hor, P. Roushan, H. Beidenkopf, J. Seo, D. Qu, J. G. Checkelsky, L. A. Wray, D. Hsieh, Y. Xia, S.-Y. Xu et al., Development of ferromagnetism in the doped topological insulator $Bi_{2-x}Mn_xTe_3$, Phys. Rev. B **81**, 195203 (2010).

[7] A. Wolos, A. Drabinska, J. Borysiuk, K. Sobczak, M. Kaminska, A. Hruban, S. G. Strzelecka, A. Materna, M. Piersa, M. Romaniec et al., High-spin configuration of Mn in $Bi_2Se_3$ three-dimensional topological insulator, J. Magn. Magn. Mater **419**, 301 (2016).

[8] J. Růžička, O. Caha, V. Holý, H. Steiner, V. Volobuiev, A. Ney, G. Bauer, T. Duchoň, K. Veltruská, I. Khalakhan et al., Structural and electronic properties of manganese-doped $Bi_2Te_3$ epitaxial layers, New J. Phys. **17**, 013028 (2015).

[9] M. M. Otrokov, T. V. Menshchikova, I. P. Rusinov, M. G. Vergniory, V. M. Kuznetsov, and E. V. Chulkov, Magnetic extension as an efficient method for realizing the quantum anomalous hall state in topological insulators, JETP Lett. **105**, 297 (2017).

[10] J. Li, Y. Li, S. Du, Z. Wang, B. L. Gu, S. C. Zhang, K. He, W. Duan, and Y. Xu, Intrinsic magnetic topological insulators in van der Waals layered $MnBi_2Te_4$-family materials, Sci. Adv. **5**, eaaw5685 (2019).

[11] N. H. Jo, L. L. Wang, R. J. Slager, J. Yan, Y. Wu, K. Lee, B. Schrunk, A. Vishwanath, and A. Kaminski, Intrinsic axion insulating behavior in antiferromagnetic $MnBi_6Te_{10}$, Phys. Rev. B **102**, 045130 (2020).

[12] C. Hu, L. Ding, K. N. Gordon, B. Ghosh, H.-J. Tien, H. Li, A. G. Linn, S.-W. Lien, C.-Y. Huang, S. Mackey et al., Realization of an intrinsic ferromagnetic topological state in $MnBi_8Te_{13}$, Sci. Adv. **6**, eaba4275 (2020).

[13] Y. Deng, Y. Yu, M. Z. Shi, Z. Guo, Z. Xu, J. Wang, X. H. Chen, and Y. Zhang, Quantum anomalous Hall effect in intrinsic magnetic topological insulator $MnBi_2Te_4$, Science **367**, aax8156 (2020).

[14] H. Deng, Z. Chen, A. Wołoś, M. Konczykowski, K. Sobczak, J. Sitnicka, I. V. Fedorchenko, J. Borysiuk, T. Heider, Ł. Pluciński et al., High-temperature quantum anomalous Hall regime in a $MnBi_2Te_4/Bi_2Te_3$ superlattice, Nat. Phys. **17**, 36 (2020).

[15] M. Gu, J. Li, H. Sun, Y. Zhao, C. Liu, J. Liu, H. Lu, and Q. Liu, Spectral signatures of the surface anomalous Hall effect in magnetic axion insulators, Nat. Commun. **12**, 3524 (2021).

[16] Y. Hu, L. Xu, M. Shi, A. Luo, S. Peng, Z. Y. Wang, J. J. Ying, T. Wu, Z. K. Liu, C. F. Zhang, et al., Universal gapless Dirac cone and tunable topological states in $(MnBi_2Te_4)_m(Bi_2Te_3)_n$ heterostructures, Phys. Rev. B **101**, 161113(R) (2020).

[17] I. I. Klimovskikh, M. M. Otrokov, D. Estyunin, S. V. Eremeev, S. O. Filnov, A. Koroleva, E. Shevchenko, V. Voroshnin, A. G. Rybkin, I. P. Rusinov et al., Tunable 3D/2D magnetism in the $(MnBi_2Te_4)(Bi2Te_3)_m$ topological insulators family, Npj Quantum Mater. **5**, 54(2020).

[18] J. Wu, F. Liu, C. Liu, Y. Wang, C. Li, Y. Lu, S. Matsuishi, and H. Hosono, Toward 2D Magnets in the $(MnBi_2Te_4)(Bi_2Te_3)_n$ Bulk Crystal, Adv. Mater. **32**, 202001815 (2020).

[19] J. Q. Yan, S. Okamoto, M. A. McGuire, A. F. May, R. J. McQueeney, and B. C. Sales, Evolution of structural, magnetic, and transport properties in $MnBi_{2-x}Sb_xTe_4$, Phys. Rev. B **100**, 104409 (2019).

[20] H. Xie, F. Fei, F. Fang, B. Chen, J. Guo, Y. Du, W. Qi, Y. Pei, T. Wang, M. Naveed et al., Charge carrier mediation and ferromagnetism induced in $MnBi_6Te_{10}$ magnetic topological insulators by antimony doping, J. Phys. D. Appl. Phys. **55**, 104002 (2022).

[21] B. Chen, F. Fei, D. Zhang, B. Zhang, W. Liu, S. Zhang, P. Wang, B. Wei, Y. Zhang, Z. Zuo et al., Intrinsic magnetic topological insulator phases in the Sb doped $MnBi_2Te_4$ bulks and thin flakes, Nat. Commun. **10**, 4469 (2019).

[22] J. Sitnicka, K. Park, P. Skupiński, K. Grasza, A. Reszka, K. Sobczak, J. Borysiuk, Z. Adamus, M. Tokarczyk, A. Avdonin et al., Systemic consequences of disorder in magnetically self-organized topological $MnBi_2Te_4/(Bi_2Te_3)_n$ superlattices, 2D Mater. **9**, 015026 (2022).

[23] C. Yan, Y. Zhu, S. Fernandez-Mulligan, E. Green, R. Mei, B. Yan, C. Liu, Z. Mao, and S. Yang, Delicate Ferromagnetism in MnBi$_6$Te$_{10}$, arXiv:2107.08137.

[24] Y. Liu, L.-L. Wang, Q. Zheng, Z. Huang, X. Wang, M. Chi, Y. Wu, B. C. Chakoumakos, M. A. McGuire, Brian C. Sales et al., Site Mixing for Engineering Magnetic Topological Insulators, Phys. Rev. X **11**, 021033 (2021).

[25] C. Śliwa, C. Autieri, J. A. Majewski, and T. Dietl, Superexchange dominates in magnetic topological insulators, Phys. Rev. B **104**, L220404 (2021).

[26] M. D. Watson, L. J Collins-McIntyre, L. R. Shelford, A. I. Coldea, D. Prabhakaran, S. C. Speller, T. Mousavi, C. R. M. Grovenor, Z. Salman, S. R. Giblin et al., Study of the structural, electric and magnetic properties of Mn-doped Bi$_2$Te$_3$ single crystals, New J. Phys. **15**, 103016 (2013).

[27] L. Zhao, M. Konczykowski, H. Deng, I. Korzhovska, M. Begliarbekov, Z. Chen, E. Papalazarou, M. Marsi, L. Perfetti, A. Hruban et al., Stable topological insulators achieved using high energy electron beams, Nat. Commun. **7**, 10957 (2016).

[28] L. Khalil, E. Papalazarou, M. Caputo, N. Nilforoushan, L. Perfetti, A. Taleb-Ibrahimi, M. Konczykowski, A. Hruban, A. Wołoś, L. Krusin-Elbaum et al., Bulk defects and surface state dynamics in topological insulators: The effects of electron beam irradiation on the ultrafast relaxation of Dirac fermions in Bi$_2$Te$_3$, J. Appl. Phys. **125**, 025103 (2019).

[29] A. Wolos, S. Szyszko, A. Drabinska, M. Kaminska, S. G. Strzelecka, A. Hruban, A. Materna, and M. Piersa, Landau-Level Spectroscopy of Relativistic Fermions with Low Fermi Velocity in the Bi$_2$Te$_3$ Three-Dimensional Topological Insulator, Phys. Rev. Lett. **109**, 247604 (2012).

[30] J. G. Analytis, J. H. Chu, Y. Chen, F. Corredor, R. D. McDonald, Z. X. Shen, and I. R. Fisher, Bulk Fermi surface coexistence with Dirac surface state in Bi$_2$Se$_3$: A comparison of photoemission and Shubnikov–de Haas measurements, Phys. Rev. B **81**, 205407 (2010).

[31] E. Papalazarou, L. Khalil, M. Caputo, L. Perfetti, N. Nilforoushan, H. Deng, Z. Chen, S. Zhao, A. Taleb-Ibrahimi, M. Konczykowski et al., Unraveling the Dirac fermion dynamics of the bulk-insulating topological system Bi$_2$Te$_2$Se, Phys. Rev. Mater. **2**, 104202 (2018).

[32] I. S. Maksymov and M. Kostylev, Broadband stripline ferromagnetic resonance spectroscopy of ferromagnetic films, multilayers and nanostructures, Phys. E **69**, 253 (2015).

[33] S.-K. Bac, K. Koller, F. Lux, J. Wang, L. Riney, K. Borisiak, W. Powers, M. Zhukovskyi, T. Orlova, M. Dobrowolska, et al., Topological response of the anomalous Hall effect in MnBi$_2$Te$_4$ due to magnetic canting, Npj Quantum Mater. **7**, 46 (2022).

[34] P. M. Sass, J. Kim, D. Vanderbilt, J. Yan, and W. Wu, Robust *A*-Type Order and Spin-Flop Transition on the Surface of the Antiferromagnetic Topological Insulator MnBi$_2$Te$_4$, Phys. Rev. Lett. **125**, 037201 (2020).

[35] N. Liu, J. Teng, and Y. Li, Two-component anomalous Hall effect in a magnetically doped topological insulator, Nat. Commun. **9**, 1282 (2018).

[36] N. Nagaosa, J. Sinova, S. Onoda, A. H. MacDonald, and N. P. Ong, Anomalous Hall effect, Rev. Mod. Phys. **82**, 1539 (2010).

[37] M. Onoda and N. Nagaosa, Topological Nature of Anomalous Hall Effect in Ferromagnets, J. Phys. Soc. Jpn. **71**, 19 (2002).

[38] D. Xiao, M. C. Chang, and Q. Niu, Berry phase effects on electronic properties, Rev. Mod. Phys. **82**, 1959 (2010).

[39] A. Wolos, M. Piersa, G. Strzelecka, K. P. Korona, A. Hruban, and M. Kaminska, Mn configuration in III-V semiconductors and its influence on electric transport and semiconductor magnetism, Phys. Status Solidi C **6**, 2769 (2009).

[40] T. Miyasato, N. Abe, T. Fujii, A. Asamitsu, S. Onoda, Y. Onose, N. Nagaosa, and Y. Tokura, Crossover Behavior of the Anomalous Hall Effect and Anomalous Nernst Effect in Itinerant Ferromagnets, Phys. Rev. Lett. **99**, 086602 (2007).

[41] H. Köhler, Non-Parabolic *E(k)* Relation of the Lowest Conduction Band in Bi$_2$Te$_3$, Phys. Stat. Sol. **73**, 95 (1976).

[42] H. Köhler, Non-Parabolicity of the Highest Valence Band of Bi$_2$Te$_3$ from Shubnikov-de Haas Effect, Phys. Stat. Sol. **74**, 591 (1976).

[43] Y. L. Chen, J. G. Analytis, J.-H. Chu, Z. K. Liu, S.-K. Mo, X. L. Qi, H. J. Zhang, D. H. Lu, X. Dai, Z. Fang et al., Experimental Realization of a Three-Dimensional Topological Insulator, Bi$_2$Te$_3$, Science **325**, 5937 (2009).

[44] B. Heinrich, Ultrathin Magnetic Structures II, ed B. Heinrich and J. A. C. Bland, "Berlin: Springer 2005" Ferromagnetic Resonance in Ultrathin Film Structures 195–222.

[45] J. Lindner and K. Baberschke, Ferromagnetic resonance in coupled ultrathin films, J. Phys. Condens. Mat. **15**, S465 (2003).

[46] B. Heinrich and J. F. Cochran, Ultrathin metallic magnetic films: magnetic anisotropies and exchange interactions, Adv. Phys. **42**, 523 (1993).

[47] Z. Zhang, L. Zhou, P. E. Wigen, and K. Ounadjela, Angular dependence of ferromagnetic resonance in exchange-coupled Co/Ru/Co trilayer structures, Phys. Rev. B **50**, 6094 (1994).